\documentclass[usegraphicx]{mn2e}

\usepackage{times}
\usepackage{amssymb}
\usepackage{amsmath}

\begin{document}

\include{defn}
\def\cm{{\rm\thinspace cm}}
\def\gm{{\rm\thinspace gm}}
\def\dyn{{\rm\thinspace dyn}}
\def\erg{{\rm\thinspace erg}}
\def\eV{{\rm\thinspace eV}}
\def\MeV{{\rm\thinspace MeV}}
\def\g{{\rm\thinspace g}}
\def\ga{{\rm\thinspace gauss}}
\def\K{{\rm\thinspace K}}
\def\keV{{\rm\thinspace keV}}
\def\km{{\rm\thinspace km}}
\def\kpc{{\rm\thinspace kpc}}
\def\Lsun{\hbox{$\rm\thinspace L_{\odot}$}}
\def\m{{\rm\thinspace m}}
\def\Mpc{{\rm\thinspace Mpc}}
\def\Msun{\hbox{$\rm\thinspace M_{\odot}$}}
\def\Zsun{\hbox{$\rm\thinspace Z_{\odot}$}}
\def\pc{{\rm\thinspace pc}}
\def\ph{{\rm\thinspace ph}}
\def\s{{\rm\thinspace s}}
\def\yr{{\rm\thinspace yr}}
\def\sr{{\rm\thinspace sr}}
\def\Hz{{\rm\thinspace Hz}}
\def\MHz{{\rm\thinspace MHz}}
\def\GHz{{\rm\thinspace GHz}}
\def\chisq{\hbox{$\chi^2$}}
\def\delchi{\hbox{$\Delta\chi$}}
\def\cmps{\hbox{$\cm\s^{-1}\,$}}
\def\cmpssq{\hbox{$\cm\s^{-2}\,$}}
\def\cmsq{\hbox{$\cm^2\,$}}
\def\cmcu{\hbox{$\cm^3\,$}}
\def\pcmcu{\hbox{$\cm^{-3}\,$}}
\def\pcmcuK{\hbox{$\cm^{-3}\K\,$}}
\def\dynpcmsq{\hbox{$\dyn\cm^{-2}\,$}}
\def\ergcmcups{\hbox{$\erg\cm^3\ps\,$}}
\def\ergpcmps{\hbox{$\erg\cm^{-3}\s^{-1}\,$}}
\def\ergpcmsqps{\hbox{$\erg\cm^{-2}\s^{-1}\,$}}
\def\ergpcmsqpspA{\hbox{$\erg\cm^{-2}\s^{-1}$\AA$^{-1}\,$}}
\def\ergpcmsqpspsr{\hbox{$\erg\cm^{-2}\s^{-1}\sr^{-1}\,$}}
\def\ergpcmcups{\hbox{$\erg\cm^{-3}\s^{-1}\,$}}
\def\ergpcmps{\hbox{$\erg\cm^{-1}\s^{-1}$}}
\def\ergps{\hbox{$\erg\s^{-1}\,$}}
\def\ergpspmp{\hbox{$\erg\s^{-1}\Mpc^{-3}\,$}}
\def\gpcm{\hbox{$\g\cm^{-3}\,$}}
\def\gpcmps{\hbox{$\g\cm^{-3}\s^{-1}\,$}}
\def\gps{\hbox{$\g\s^{-1}\,$}}
\def\Jy{{\rm Jy}}
\def\keVpcmsqpspsr{\hbox{$\keV\cm^{-2}\s^{-1}\sr^{-1}\,$}}
\def\kmps{\hbox{$\km\s^{-1}\,$}}
\def\kmpspmp{\hbox{$\km\s^{-1}\Mpc{-1}\,$}}
\def\Lsunppc{\hbox{$\Lsun\pc^{-3}\,$}}
\def\Msunpc{\hbox{$\Msun\pc^{-3}\,$}}
\def\Msunpkpc{\hbox{$\Msun\kpc^{-1}\,$}}
\def\Msunppc{\hbox{$\Msun\pc^{-3}\,$}}
\def\Msunppcpyr{\hbox{$\Msun\pc^{-3}\yr^{-1}\,$}}
\def\Msunpyr{\hbox{$\Msun\yr^{-1}\,$}}
\def\pcm{\hbox{$\cm^{-3}\,$}}
\def\pcmsq{\hbox{$\cm^{-2}\,$}}
\def\pcmK{\hbox{$\cm^{-3}\K$}}
\def\phpcmsqps{\hbox{$\ph\cm^{-2}\s^{-1}\,$}}
\def\pHz{\hbox{$\Hz^{-1}\,$}}
\def\pmpc{\hbox{$\Mpc^{-1}\,$}}
\def\pmpccu{\hbox{$\Mpc^{-3}\,$}}
\def\ps{\hbox{$\s^{-1}\,$}}
\def\psqcm{\hbox{$\cm^{-2}\,$}}
\def\psr{\hbox{$\sr^{-1}\,$}}
\def\kmpspMpc{\hbox{$\kmps\Mpc^{-1}$}}

\voffset=-0.4in

\title{Radiative pressure feedback by a quasar in a galactic bulge}
\author[A.C. Fabian, A. Celotti, M.C. Erlund]
{\parbox[]{6.in} {A.C. Fabian$^1$, A.~Celotti$^2$ and M.C. Erlund$^1$\\
\footnotesize
$^1$Institute of Astronomy, Madingley Road, Cambridge CB3 0HA\\
$^2$S.I.S.S.A., via Beirut 2-4, 34014, Trieste, Italy\\}}

\maketitle 

\begin{abstract} We show that Eddington-limited black hole
luminosities can be sufficient to deplete a galaxy bulge of gas
through radiation pressure, when the ionization state of the gas and
the presence of dust are properly taken into account. Once feedback
starts to be effective it can consistently drive all the gas out of
the whole galaxy.  We estimate the amount by which the effect of
radiation pressure on dusty gas boosts the mass involved in the
Eddington limit and discuss the expected column density at which the
gas is ejected.  An example is shown of the predicted observed nuclear
spectrum of the system at the end of an early, obscured phase of
growth when the remaining column density $N_{\rm H}\sim 
10^{24}f\pcmsq$, where $f$ is the gas fraction in the bulge.  

\end{abstract} 

\section{Introduction}

Much observational work over the past decade has shown that the mass
of the central black hole in a galaxy, $M_{\rm BH}$, scales with the
mass and/or velocity dispersion, $\sigma$, of the bulge of that galaxy
(Kormendy \& Richstone 1995; Magorrian et al 1998; Gebhardt et al
2000; Ferrarese et al 2001).  A recent correlation (Tremaine et
al 2002) shows that $M_{\rm BH}\propto \sigma^4$ holds over at least 3
decades in black hole mass. This result suggests that black hole
and galaxy growth are entwined and introduces the exciting possibility
that the growth of a central black hole determines the properties of
its host galaxy.

Many models have been produced to explain the $M_{\rm BH} - \sigma$
correlation. The ones relevant here involve a central active galaxy
influencing the level of gas, and thus star formation, of the host
galaxy. The black hole may for example grow in mass and power until it
is capable of ejecting the interstellar medium from the galaxy, thus
stopping star formaton and determining the total stellar mass of the
galaxy. Such models either employ an energy argument (Silk \& Rees
1998; Haehnelt, Natarajan \& Rees 1998; Wyithe \& Loeb 2003) which
leads to $M_{\rm BH}\propto \sigma^5$ or a momentum one using a quasar
wind (Fabian 1999) or the quasar radiation directly (Fabian et al
2002; King 2003; Murray et al 2004), to obtain $M_{\rm BH}\propto
\sigma^4$. More general heating models have been
presented by Granato et al (2005) and by Sazanov et al (2005), while
Begelman \& Nath (2005) explored the role of momentum in
self-regulating the gas density profile.

The binding energy of the likely interstellar medium of a galaxy is
less than one per cent of the energy released by the growth of its
massive central black hole. This energy must of course be supplied in order to
eject the gas. Sufficient momentum is also essential, which
makes the overall process energetically inefficient.


Here we revisit the momentum approach using radiation pressure from a
central quasar acting on the dense star-forming interstellar medium of
a young galaxy. The obvious approach is to use the Eddington limit,
but in its original form it applies to a point mass while here we wish
to deal with a galaxy which is a distributed mass. If the standard
Eddington limit cannot be exceeded and is applied to both quasar and
galaxy, then there is no way that the quasar radiation can eject the
surrounding galactic medium since the mass and thus the
Eddington-limiting luminosity required rises with radius.  One
approach (King 2003) is to invoke super-Eddington radiation levels
from the quasar. There is, however, little observational evidence for
super-Eddington radiation (e.g. Woo \& Urry 2002; Kollmeier et al
2005) or firm theoretical basis for it (but see Begelman 2002). Here
we introduce an {\em effective} Eddington limit which occurs when
radiation acts on lowly ionized and dusty gas. The quasar can easily
exceed this limit and so drive gas from the galaxy. Of particular
interest here is the column density of the gas within the galactic
bulge.

\section{The Eddington Limit} 

We now derive an effective Eddington limit for the situation when the
radiation from a central active galaxy or quasar interacts with dusty,
partially-ionized gas. The effective interaction cross-section, due to
photoelectric absorption, dust extinction etc. can then be much larger
than the Thomson cross-section used to derive the standard Eddington
limit. A central quasar at the standard Eddington limit for its mass,
relevant for highly ionized gas in its immediate vicinity, can be
radiating at, or above. the {\em effective} Eddington limit for
distant matter gravitationally bound by the higher mass of the black
hole and the host galaxy.

The Eddington limit arises when the outgoing radiation pressure, due
to electron scattering, from a source of luminosity $L$ balances the
gravitational attraction due to its mass $M$: 
\begin{equation} 
L_{\rm Edd}={{4 \pi G M m_{\rm p} c}\over {\sigma_{\rm T}}}.
\end{equation} 

It is assumed here that radiation pressure is acting on
a gas of ionized hydrogen around the photon source. $G, m_{\rm p}$ and
$\sigma_{\rm T}$ are the gravitational constant, proton mass and
Thomson cross-section, respectively. If the cross-section for
interaction between the radiation and matter $\sigma_{\rm i}$ is
larger than for electron scattering, then the relevant limit which we
denote the effective Eddington limit $L^{\prime}_{Edd}$ is
proportionally changed.  Furthermore, the Eddington limit is often derived 
by considering an isolated
electron-proton pair  exposed to the radiation force. Here we need the
effects of radiation on shells of matter surrounding the source so we
consider the effective limit for a column density of gas $N$.
For a gas optically thin to Thomson scattering, 
$L^{\prime}_{\rm Edd}\simeq L_{\rm Edd} \tau_{\rm
  T}/{{\rm min}[\tau_{\rm i}, 1]}$, where $\tau_{\rm T}$ and  $\tau_{\rm i}$ 
are the optical depths for the corresponding cross-sections,  
$\tau \equiv \sigma N$. 
 
This means that a luminosity which is sub-Eddington for completely
ionized gas close to a central mass, $M$, can exceed the modified
Eddington limit for partially-ionized or neutral gas, for which
$\sigma_{\rm i} > \sigma_{\rm T}$, that is denser or further away.
Note that $\sigma_{\rm i}$ is an effective cross-section obtained by
averaging over the incident spectrum, the column density of matter and
the state of that matter (ionization state, dust content, chemical
composition, etc.). It therefore depends on the spectral shape of the
incident radiation. Here we refer to the integrated spectrum and thus
to an average cross-section.  The absorbed luminosity correponds to
$L_{\rm a}\simeq L \tau_{\rm i}$ in the optically thin regime and
$L_{\rm a} \simeq L$ for optically thick gas. We assume $L_{\rm a}$ is
radiated isotropically by the absorber and thus resultant rate of
change of momentum per unit area, or radiation pressure, is $L_{\rm
  a}/4\pi r^2 c$.

The amplification factor $A$ owed to the presence of gas not fully
ionized and dust, can be defined as the ratio of the effective
radiation pressure $L {\rm min}[\tau_{\rm i}, 1]/4\pi r^2 c $ acting
outward on a column $N$ gas at radius $r$ with respect to that for fully ionized gas. 
This corresponds to
\begin{equation} A = {L_{\rm a} \over L \tau_{\rm T}} = {{\rm min}[\tau_{\rm i}, 1]\over \tau_{\rm T}}.
\end{equation}

Alternatively, this can be re-expressed in terms of the mass required
to gravitationally hold this column density back from expulsion, which
in terms of $A$ can be written as
\begin{equation} 
M^{\prime}_{\rm Edd}= A M_{\rm BH}.
\end{equation} 
The latter expression  assumes that the central black hole, of
mass $M_{\rm BH}$, is radiating at its (Thomson scattering) Eddington
limit, so $L=L_{\rm Edd}$.

\section{The Eddington Boost Factor}

We have determined the boost factor, $A$, by finding the luminosity
absorbed by column density, $N_{\rm H}$, after making assumptions
about the ionization parameter of the gas and incident radiation
spectrum. For this purpose we use the code {\sc cloudy} 96.01 (Ferland
1993) with the AGN spectrum. This essentially assumes a UV blackbody
plus an X--ray powerlaw spectrum appropriate for a supermassive,
radiatively-efficent, accreting black hole.
 
The boost factor is obtained from the directly absorbed radiation. We
subtract from the input luminosity that which is transmitted (without
any diffuse radiation which is assumed to be isotropic). Basically we
are using the continuum radiation pressure and assuming that the
absorbed radiation emerges as heat which, on emission, interacts
either little or with no dynamical consequences on the galactic gas.
Trapping of radiation is assumed to be negligible. The procedure has
been stepped over a range of column density $N_{\rm H}$ and ionization
parameter $\xi=L/nr^2$ with the results plotted as $A$ in Fig.~1a.
Different symbols represent the cases with no dust and gas with a
Galactic mix of dust. We expect that this covers the conditions in
growing galaxies. If there is vigorous star formation and a high
metallicity in the gas surrounding growing massive black holes then
the gas should be very dusty. 

Overall the boosting factor is independent of $N_{\rm H}$ (see
particularly the cases without dust), and just given by the ratio of
the effective optical depth with respect to the Thomson one, until the
gas becomes optically thick.  For larger columns the radiation
pressure does not increase further, while gravity acts on a more
massive shell, leading to $A\propto N^{-1}$.  The normalization of $A$
in the optically thin regime and the column corresponding to an
effective depth of unity clearly depends (inversely and directly,
respectively) on the ionization state of the gas, parametrized by
$\xi$. 

As shown below, the effect of dust is to boost the effect of radiation
pressure by one or more orders of magnitude, which -- as discussed
later -- is key to ensure that gas can be depleted from a proto-galaxy
for Eddington luminosities from the central black hole. The
transmitted and total spectra for a column density $N_{\rm
H}=10^{23}\psqcm$ and ionization parameter $\xi=100$ is shown in
Fig.~2 for both the dusty and dust-free cases. It is clear that much
more radiation is absorbed in the dusty case with the radiation from
the nucleus much reduced at energies above 0.3~eV or shorted than 3.6
microns wavelength. This means that the radiation pressure from much
of the large UV/optical blackbody emission is harnessed so causing the
threefold or more greater boost factor $A,$ when compared with the
dust-free case. Since most of the radiation is absorbed in the dusty
case ($L_{\rm a}$ is close to $L$), Fig.~1 follows straightforwardly
from equation 2. The precise energy of the resulting infrared emission
bump depends on the dust temperature and thus on the radial
distribution of the dust.

The energy of the blackbody emission, assumed to originate from an
accretion disc around the black hole, depends on the black hole mass,
shifting to higher energies as the mass reduces. This means that the
difference between the dusty and dust-free cases reduces for lower
mass objects ($M_{\rm BH}<10^8\Msun$). The blackbody peak is absorbed
for all relevant masses for the dusty case, so we expect the resulting
$M-\sigma$ relation to be robust where the other assumptions hold. 

\begin{figure} 
\includegraphics[width=\columnwidth]{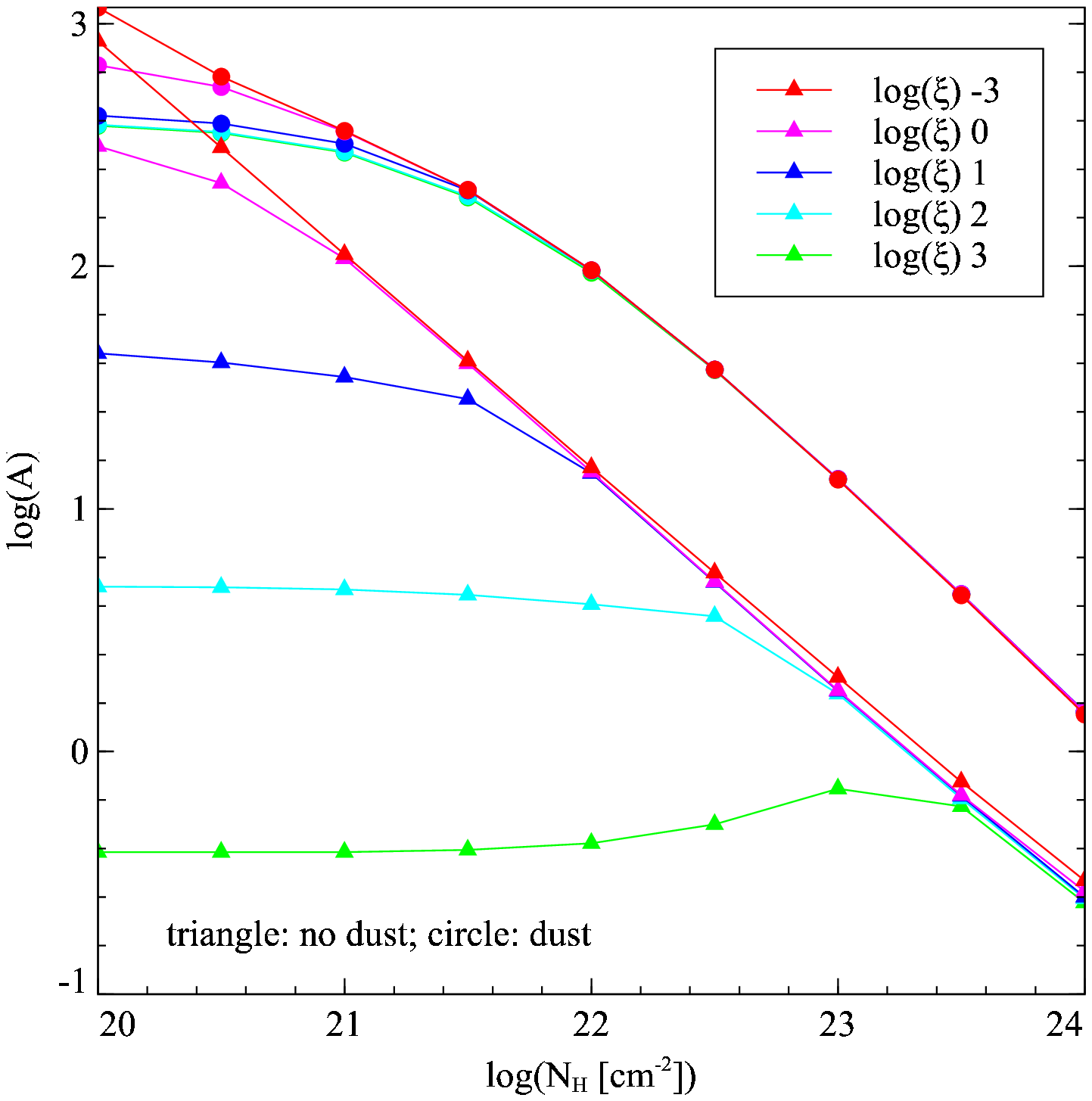}
\includegraphics[width=\columnwidth]{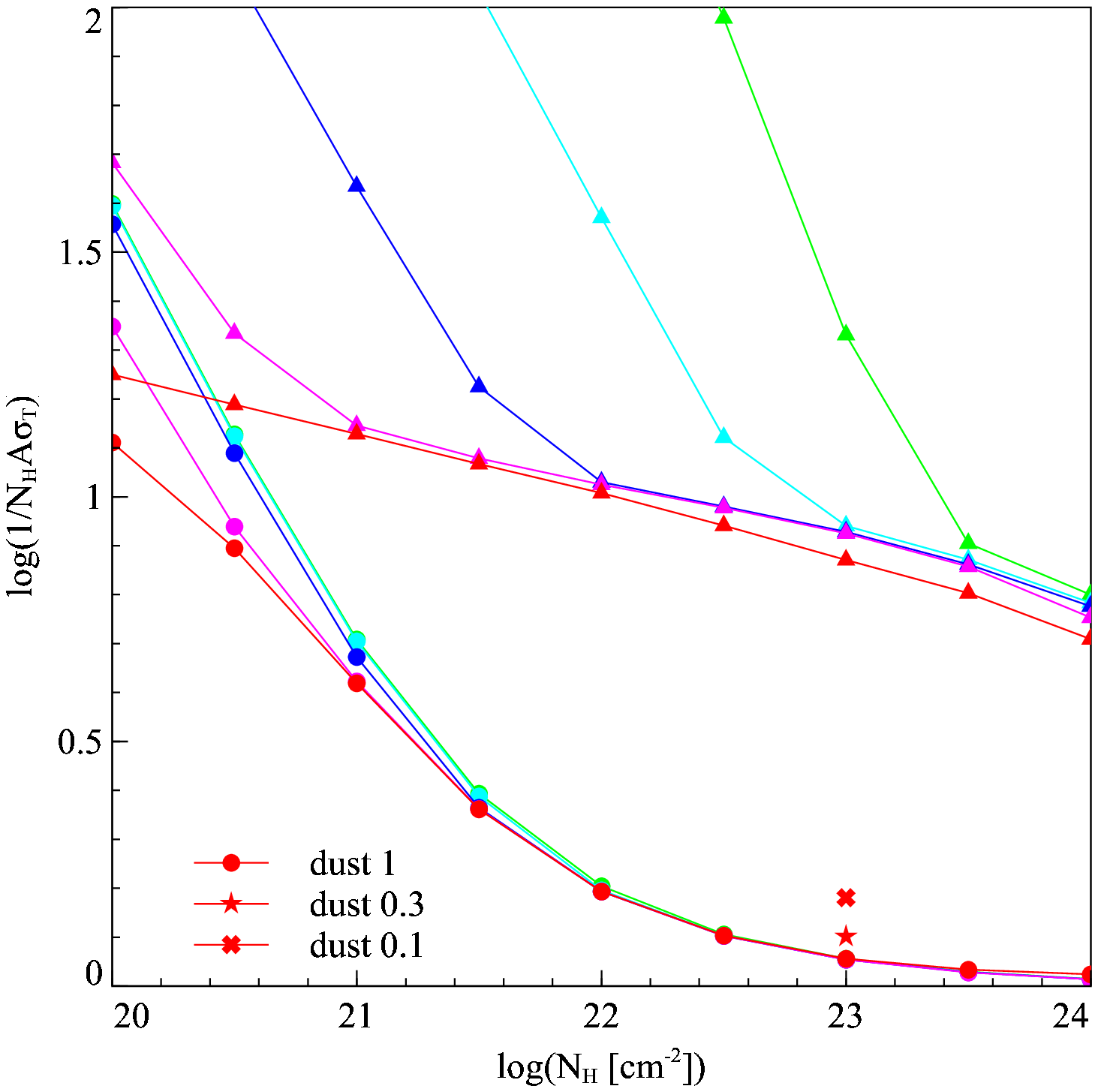} 
\caption{a) Top:
Boost factor $A$ as a function of column density (expressed in terms
of neutral hydrogen column $N_{\rm H}$) and for varius values of the
ionization parameter $\log\xi$ from -3 to 3. Circles denote gas with a
Galactic mix of dust. b) Bottom: Factor $1/N_{\rm H} A \sigma_{\rm T} =
L/L_{\rm a}$ shown as a function of column density. For dusty ionized
gas, the factor is close to unity for $N_{\rm H}>10^{23}\psqcm$ and
less than 2 for $N_{\rm H}>5\times 10^{21}\psqcm$. The effect of
having the dust-to-gas ratio at 0.3 and 0.1 times the Galactic ISM
value are shown for a column density of $10^{23}\psqcm.$  } 
\end{figure}


\section{Expulsion of gas from a galaxy}

Let us consider now the specific case of a galactic bulge, and in particular follow the 
scenario discussed by Fabian (1999) and Fabian et al. (2002). It is assumed that the bulge 
is isothermal with mass-density
profile $\rho\propto r^{-2}$. The total mass within radius $r$  is given by
\begin{equation}
M = {{2 \sigma^2 r}\over G}
\end{equation}
of which fraction $f$ is assumed to be in gas, and this has been 
accreted onto the black hole, so $M_{\rm BH}=fM.$
The column density exterior to $r$ is then
\begin{equation}
  N={{f\sigma^2}\over {2\pi G m_{\rm p} r}}.
\end{equation} 

It should be noted that in Section 2, we treated the gravitational
force as acting on gas concentrated in a thin shell at distance $r$,
while the bulge gas is distributed, and this leads to an extra factor
${\ln} (r_{\rm max}/r)$ where $r_{\rm max}$ corresponds to the outer
boundary of the isothermal distribution (Fabian et al 2002).

As previously mentioned, accretion onto the black hole within $r$
leaves a column density given by equation (5) beyond.  Our treatment
of such a column being concentrated in a shell corresponds to the
possible scenario where the very same radiation pressure would
compress the gas into a shell propagating outward.  Alternatively, one
has to consider the extra logarithmic factor: although the value of
$r_{\rm max}$ is not clearly determined a priori, there has to be a
outer boundary of the (otherwise diverging) isothermal mass
distribution $\propto r^{-2}$. Finally, we note that eqn. (5) also
arises if the gas within $r$ is assumed to be swept up into a shell
rather than accreted into the black hole.

The question we want to quantitatively answer here is whether 
radiative feedback from the central accreting black hole is sufficient to 
deplete gas on the galactic scale.  

Gas is expelled from the bulge when the luminosity of the accreting
black hole $L$ exceeds the modified Eddington luminosity, namely
$A>1$.  Or more precisely, gas at radius $r$ is pushed outward when
$M^{\prime}_{\rm Edd}$ for the column external to the gas exceeds the
mass internal to $r$. Combining eqs. (2), (4), (5) with $L=L_{\rm
  Edd}$, this corresponds to
\begin{equation}
{L _{\rm a}\over{4\pi c}}= {{f\sigma^4}\over{\pi G }},
\end{equation}
or   from eqs. (3), (4) and (5)
\begin{equation}
M_{\rm BH}={{f\sigma^4}\over{\pi G^2 m_{\rm p} AN}}.
\end{equation} 

The computation presented in Section 3 shows (Fig.~1b) that
$AN\sigma_{\rm T}$ is close to unity over a wide range of column
densities from $10^{22}$ to $10^{24}\pcmsq$. This means that we can
replace $AN$ by ${\sigma_{\rm T}}^{-1}$ in the above equation so
obtaining \begin{equation} M_{\rm BH}={{f\sigma^4 }\over{\pi G^2
      m_{\rm p}}}\sigma_{\rm T}.
\end{equation}

This is the same expression, within a factor of two, as that derived on
simpler grounds by Fabian (1999), Fabian et al (2002), King (2003) and
Murray et al (2004). What we have done here is to consider a more
realistic gas composition and incident spectrum, and to  
quantitatively consider the effect of dust. The result is
obtained principally because most of the relevant incident radiation,
that in the UV to soft X-ray bands, is absorbed by column densities
greater than $10^{22}\pcmsq$ of dusty gas which is not completely
ionized. Note that the resulting black hole mass (eqn 8) depends on
the Eddington fraction of the source as $f_{\rm Edd}^{-1}$, so is
larger for sources operating below the Eddington limit. Kollmeier et
al (2005) find in their sample of quasars at redshifts $z=0.3-2$ that
most are within a factor of ten of the Eddington limit.

We have assumed that the outer gas is all in a shell at radius $r$,
which slightly overestimates the force required. Radiation pressure
will sweep matter into a dense shell, the column density of which will
evolve as $N\propto r^{-1}$. Provided that it is still within the
regime where $AN$ is constant, then it will still be driven outward. 

The ionization parameter of matter at radius $r$ when it is about to
be ejected is 
\begin{equation}
\xi={{L 2\pi G m_{\rm p}}\over{f \sigma^2}}.
\end{equation}
but at that point $L\approx L_{\rm a}$, so from (6)
\begin{equation}
\xi=8\pi c m_{\rm p}\sigma^2=160\sigma_2^2,
\end{equation}
where we use $\sigma=100\sigma_2\kmps.$ This means that the line in
Fig.~1 for $\xi=100$ is the important one for the general case. We
envisage that the gas is dusty and so the upper lines in Fig.~1a
are most relevant. The input and output spectrum for such a case is
shown in Fig.~2.  

\begin{figure}
\includegraphics[width=\columnwidth]{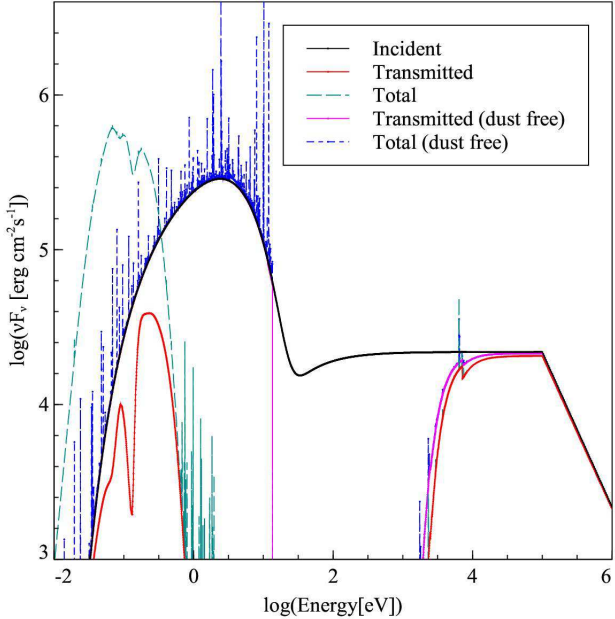}
\caption{Input, transmitted and total emergent spectra for a
dusty
shell with column density $N_{\rm H}=10^{23}\psqcm$ and ionization
parameter $\xi=100$. The corresponding spectra for no dust are also
shown. The normalization of the y-axis is arbitrary.}
\end{figure}

\section{Discussion}

In a very simple model, we envisage that the black hole grows by
accreting the inner gas. Its mass $fM(r)$ then 
  increases with time, its (standard
Eddington-limited) luminosity $L$ rises with mass and the column
density $N$ of the gas beyond $r$ decreases until the condition given
by equation (3) is met. The value of $A$ at this point is $1/f$ and
$N\approx fN_{\rm T}$. For gas fractions of 10 per cent the final
column density is then $\sim 10^{23}\psqcm.$ The gas is then ejected
from the host galaxy.

If the luminosity is sub-Eddington then $N\approx ff_{\rm Edd} 
N_{\rm T}.$ The whole mechanism will fail if $ff_{\rm Edd}$ is much
less than 0.01 at which point $AN$ is no longer close to $N_{\rm T}$. 

The inner radius of the gas when  expulsion occurs is (from eq. (5)) 
\begin{equation}
r_{\rm exp}={\sigma^2\over{2\pi G m_{\rm p}} }\sigma_{\rm T}.
\end{equation}
This is a factor $(2f)^{-1}$ times the accretion radius ($r_{\rm a}
=GM/\sigma^2$), so for a typical value of $f=0.1,$ $r_{\rm exp}\sim 5
r_{\rm a}$. The region where most of the obscuration occurs is
therefore compact. 

Most of the black hole growth will have been by obscured accretion, as
implied by the observed X-ray background (Fabian \& Iwasawa 1999; see
also Fabian 2004; Brandt \& Hasinger 2005; Worsley et al 2004;
Alexander et al 2005; Civano, Comastro \& Brusa 2005;
Martinez-Sansigre et al 2005) and mid-infrared studies with Spitzer
(Treister et al 2005). There is already a considerable
population of known luminous AGN with column densities $N_{\rm H}\sim
10^{23} - 5\times 10^{23}\psqcm$. In our model the main black hole
growth phase occurs as the column density reduces to $fN_{\rm T}.$ 

We do of course require that the black hole continues to be fuelled
for the gas expulsion time which is several tens of millions of years.
This could involve a torus and disc around the black hole, such as
found in many models for AGN (see e.g. Antonucci 1993). Anisotropy of
the radiation due to such structures will also cause the ejection to
not be spherically symmetric. We consider that such issues are
secondary to the basic model outlined above. 

We have assumed that the gas is mostly cold and yet distributed
throughout the bulge of the host galaxy in a manner similar to that of
the stars. This requires that the gas is supported in some way and we
assume that a pervasive hotter phase may be responsible. Cold clouds
embedded in a hotter medium can drag the hotter medium with them when
ejected. Why the distribution should be $\rho\propto r^{-2}$ is
unexplained, although it must roughly occur if the gas clouds form
into stars which appear to have this distribution. 

The feedback process described here ultimately switches off growth of
both black hole and host galaxy by radiation from the accreting black
hole interacting with, and expelling, surrounding cold gas.  Note that
for the feedback mechanism to be effective implicitly allows for the
cold gas component in the bulge to convert in stars.  Different
feedback processes are expected in massive galaxies, particularly
those at the centres of groups and clusters where most surrounding gas
is hot and ionized and so immune from the effects of radiation
pressure on neutral and partially ionized effects. In such massive
objects other processes are required for feedback (e.g. Churazov et al
2005) and the slope and normalization of the $M-\sigma$ relation may
change (Fabian et al 2005).

\section{Acknowledgments} We acknowledge Gary Ferland for the use of
his code {\sc cloudy} and a referee for comments. The Royal Society
(ACF), the  Italian MIUR and INAF (AC)
and PPARC (MCE) are acknowledged for financial support.
This research was supported in part by the National Science Foundation
under Grant No. PHY99-07949; the KITP (Santa Barbara)
is thanked for kind hospitality (AC).

\end{document}